# Can a Future Choice Affect a Past Measurement's Outcome?

Yakir Aharonov[1,2,3], Eliahu Cohen[1*] Avshalom C. Elitzur[3]

*An EPR experiment is studied where each particle within the entangled pair undergoes a few weak measurements (WMs) along some pre-set spin orientations, with the outcomes individually recorded. Then the particle undergoes one strong measurement along an orientation chosen at the last moment. Bell-inequality violation is expected between the two final measurements within each EPR pair. At the same time, statistical agreement is expected between these strong measurements and the earlier weak ones performed on that pair. A contradiction seemingly ensues: i) Bell's theorem forbids spin values to exist prior to the choice of the orientation measured; ii) A weak measurement is not supposed to determine the outcome of a successive strong one; and indeed iii) Almost no disentanglement is inflicted by the WMs; and yet iv) The outcomes of weak measurements statistically agree with those of the strong ones, suggesting the existence of pre-determined values, in contradiction with (i). Although the conflict can be solved by mere mitigation of the above restrictions, the most reasonable resolution seems to be that of the Two-State-Vector Formalism (TSVF), namely, that the choice of the experimenter has been encrypted within the weak measurement's outcomes, even before the experimenters themselves know what their choice will be.*

[1] School of Physics and Astronomy, Tel-Aviv University, Tel-Aviv 6997801, Israel.
[2] Schmid College of Science, Chapman University, Orange, CA 92866, USA.
[3] *Iyar*, The Israeli Institute for Advanced Research, Rehovot, Israel.
* eliahuco@post.tau.ac.il

# Introduction

Bell's theorem [1] has dealt the final blow on all hopes to explain the EPR correlations [2] as previously determined. Bell proved that these cosine-like correlations also depend on the two particular spin-orientations chosen for each measurement. As these choices can be made *at the last moment*, the resulting combinations of measurement outcomes, being mutually exclusive, *could not co-exist in advance*. Consequently, nonlocal effects between the particles have been commonly accepted as the only remaining explanation.

A variation of the EPR experiment is hereby presented, however, that suggests a simpler local explanation, namely allowing causation to be time-symmetric in the quantum realm. Then, what appears to be nonlocal in *space* turns out to be perfectly local in *spacetime*. This account's gist is given in Fig. 2.

The outline of this paper is as follows. Sec. 1 introduces the foundations of the Two-State-Vector Formalism (TSVF) and 2 weak measurement (WMs). 3 describes a combination of strong and weak measurements on a single particle illustrating a prediction of TSVF. In 4 we proceed to the EPR-Bell version of this experiment. Secs. 5-6 discuss and summarize the predicted outcomes' bearings.

## 1. A Particle's State between Two Noncommuting Measurements

Consider a particle undergoing two consecutive strong (*i.e.*, projective) measurements, along the co-planar spin orientations α and β (the strong-weak distinction will be further discussed in Sec. 2). The correlation between their outcomes depends on their relative angle $\theta_{\alpha\beta}$:

$$\langle \sigma_\alpha \sigma_\beta \rangle = \cos\theta_{\alpha\beta}. \tag{1}$$

Also, by the uncertainty relations between spin operators, these two measurements disturb each other's outcomes: If, *e.g.*, the α measurement is repeated after the β, when the two directions are orthogonal, then the initial value of the *spin-α* measurement may flip to the opposite value with probability of *1/2*.

Aharonov, Bergman and Lebowitz (ABL) [3] argued that, at any time between the two measurements, the state of the particle is equally determined by *both* backward and forward time-evolving boundary conditions. The probability for measuring the eigenvalue $c_j$ of the observable *c*, given the initial and final states $|\Psi(t')\rangle$ and $\langle\Phi(t'')|$, respectively, is described by the symmetric formula

$$P(c_j) = \frac{|\langle\Phi(t)|c_j\rangle\langle c_j|\Psi(t)\rangle|^2}{\sum_i |\langle\Phi(t)|c_i\rangle\langle c_i|\Psi(t)\rangle|^2}, \tag{2}$$

thus having a *definite* value which agrees with both outcomes due to two state-vectors, one evolving from the past

$$|\psi(t)\rangle = \exp(\int_{t'}^{t} -iH/\hbar dt)|\psi(t')\rangle, \quad t > t', \tag{3}$$

and the other from the future:

$$\langle\Phi(t)| = \langle\Phi(t'')|\exp(\int_{t''}^{t} iH/\hbar dt), \quad t < t'', \tag{4}$$

creating the two-state-vector

$$\langle\Phi(t'')| \quad |\psi(t')\rangle, \tag{5}$$

which holds for *every* intermediate moment in the evolution of the quantum system. This combination of forward and backward-evolving wave states taken from two Hilbert spaces, is argued to better describe a quantum system between two strong measurements. It is also the one which gives rise to the so called "weak value" [10-14].

TSVF accords with earlier physical models which sought to explain the spatiotemporal oddities of quantum interactions, such as [4-7]. Even the Wheeler-Feynman "absorber theory" [8] originally proposed to explain classical electromagnetic interaction, was later generalized by Cramer [9] into a comprehensive model for all quantum interactions. Admittedly, TSVF is not acknowledged as superior to more conservative, one-vector formulations of quantum mechanics. Neither has its approach to the measurement problem [15] been universally accepted. In what follows, however, we stress its rigor, elegance and simplicity, and in a consecutive papers [16,17] we present novel predictions that appear more natural with two-time vectors.

## 2. Weak measurements

TSVF, however, is unique among these models in that it has derived several predictions that, although fully consistent with the standard formalism (see Appendix 2), seem surprising and more acutely opposed to classical laws. In addition, TSVF has produced a unique technique for observing these predictions, namely WM [10-14]. For a simple and up-to-date introduction to WM's underlying principles and its rigor in cases where projective measurement proves inadequate, see [11].

In brief (taking spin measurement as an example), WM is best performed on a large ensemble of particles, weakly coupling (through a unitary interaction) each spin of the ensemble to a macroscopic device. The measuring device, has a quantum pointer (described in this case as a continuous observable $Q_d$) moving $\lambda/\sqrt{N}$ or $-\lambda/\sqrt{N}$ units upon measuring, respectively, ↑ or ↓ spin along a certain direction, where $N$ is the size of the ensemble and $\lambda$ is a constant. Let the pointer value have a Gaussian noise with 0 expectation and $\delta \gg \lambda/\sqrt{N}$ standard deviation, hence, the weakness of the measurement - the tiny bias of the quantum pointer is much below its uncertainty. When weakly measuring a single spin, we get most of the results within a wide range $\lambda/\sqrt{N} \pm \delta$. But when summing up the *N/2* weak results that correspond to the future ↑ strong results, we find most of them within a much narrower range $\lambda\sqrt{N}/2 \pm \delta\sqrt{N}/\sqrt{2}$, thereby agreeing with the strong result when $\lambda \gg \delta$.

We now move from this spin example to the general case. Let $N \gg 1$ particles undergo an interaction according to the Hamiltonian (for an alternative description using POVM see Appendix 3):

$$H_{\text{int}}(t) = \frac{\lambda}{\sqrt{N}} g(t) A_s P_d, \tag{6}$$

where $A_s$ denotes the measured observable and $P_d$ is canonically conjugated momentum to $Q_d$, representing the measuring device's pointer position. The coupling *g(t)* is nonzero only for the time interval $0 \leq t \leq T$ and normalized according to

$$\int_0^T g(t)dt = 1. \tag{7}$$

After the quantum particle weakly interacts with the quantum pointer according to Eq. 6, the pointer's movement is amplified and classically recorded. That is, the pointer undergoes a "strong measurement", but the outcome tells us very little about the individual state of any of the *N* particles.

The weakness of the measurement (and consequently its strength in preserving the wavefunction of the particles) is due to the small factor[4] $\lambda/\sqrt{N}$, inversely proportional to the square root of the ensemble's size, where $\lambda$ is a constant. When the $N$ particles have different states, e.g., spins, WM correctly gives their average. But when they all share the same ↑ or ↓ spin value along the same orientation, WM indicates the entire ensemble's state. As pointed out in [14]:

$$\langle \bar{A} \rangle_w = \frac{1}{N} \sum_{n=1}^{N} \langle A^{(n)} \rangle_w = \langle \psi | A | \psi \rangle, \tag{8}$$

where the subscript $w$ denotes weak values, and $n$ enumerates the measured particles. Eq. 8 states that the weak value of $\bar{A}$ approaches the expectation value of $A$ operating on $|\psi\rangle$.

The slicing method [11], presented below, enables isolating such same-state particles even within random ensembles. For such homogenous sub-ensembles, WM's rigor approaches that of strong measurement.

## 3. Combining Strong and Weak Measurements

We are now in a position to propose an experimental demonstration of the ABL-TSVF main argument (Fig.1), namely: A particle's state between two strong measurements carries both the past and future outcomes.

---

[4] Weakness of 1/N is sufficient in this case where one measuring apparatus is used, but for the cases considered in the next sections we chose $1/\sqrt{N}$ interaction strength. See also [14].

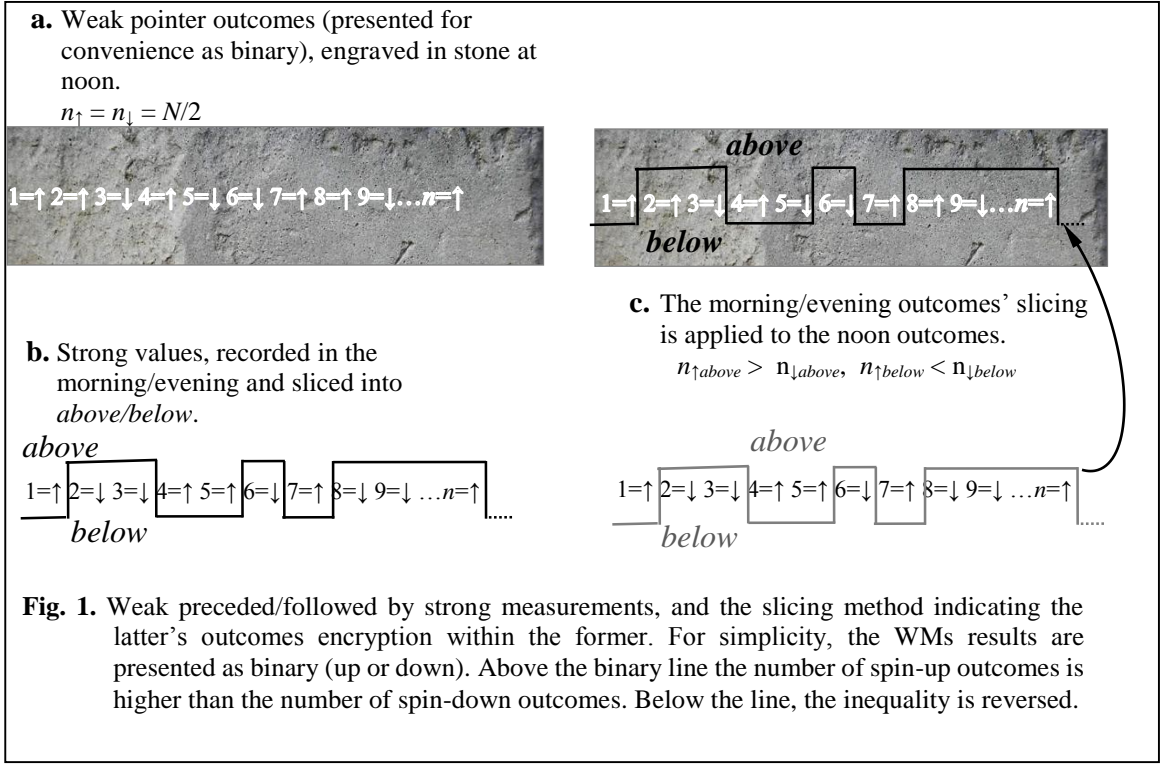

**Fig. 1.** Weak preceded/followed by strong measurements, and the slicing method indicating the latter's outcomes encryption within the former. For simplicity, the WMs results are presented as binary (up or down). Above the binary line the number of spin-up outcomes is higher than the number of spin-down outcomes. Below the line, the inequality is reversed.

Consider an ensemble of $N$ particles. Then,

### 3.1. Procedure

a) In the morning Bob[5] strongly measures all the spins along the α-orientation. He measures them one by one (a "single shot" experiment), assigning each particle a serial number such that it remains individually distinguishable throughout. He then separates his series of outcomes into two groups such that all ↑ outcomes are in one group and all ↓s are in another. Graphically, we shall denote this procedure by draws a binary line ⌐⌐⌐ dissecting Bob's row of outcomes, all ↑s left above it and all ↓s below.

b) At noon Alice weakly measures all the spins along the α orientation, as well as β, plus a third coplanar orientation γ, where α, β and γ are arbitrary non-parallel directions. For the sake of later purposes we will assume that the angles α, β and γ are those which maximize the violation of Bell's inequality. Her measurements are similarly individual, each particle measured in its turn, and the measuring device calibrated before the next measurement. To make the results more striking, she repeats this series 3 times, total 9 weak measurements per each particle. All lists of outcomes[6] are then publically recorded, *e.g.*, engraved on stone (see Fig. 1), along 9 rows with each outcome's position along the row being equal to that of Bob's list. Summing up her α-measurements (whether $α^{(1)}$, $α^{(2)}$, $α^{(3)}$ separately or all *3N* together), she finds the spin distribution to be approximately 50%↑-50%↓. Similarly for β and γ. In other words – nothing unusual.

---

[5] Along this paper, Alice and Bob are time-like separated, in contrast to the more common convention when they are space-like separated.

c) In the evening Bob, *oblivious of Alice's noon outcomes*, again strongly measures all *N* particles, this time along the β orientation. Again he draws a binary line ⌐⌙⌐ as in (a).

d) Bob then gives Alice the two binary lines, without disclosing to her whether "above/below line" refers to ↑/↓ or whether the orientation chosen for the morning/evening series was α, β or γ.

e) Alice slices her data, according to Bob's divisions. In terms of Fig. 2, she merely shifts each of Bob's ⌐⌙⌐ lines across her 9 rows of outcomes carved on stone. Each of the *N* sequences is thereby split into two approximately *N/2* "above/below line" sub-rows (to the overall of 18 subgroups). She then re-sums each half separately.

---

[6] Although each measurement's result can be any real number, for simplicity we describe Alice's results as binary - "up" for positive numbers and "down" for negative numbers.

Each row is thus sliced twice, first according to Bob's morning and then evening line.

### 3.2. Predictions

Upon Alice's re-summing up each of her sliced lists,

a) Out of the 9 sliced rows of the WM outcomes, 3 immediately stand out with maximal correlation with Bob's above/below list (see Appendix 1 for a more detailed explanation of these correlations), indicating that $x=\alpha$, above=↑, below=↓. Similarly for Bob's evening list: 3 other rows now reveal that $y=\beta$, above=↑ below=↓. In short, *the overall sum of weak measurements outcomes (after slicing) agrees with the strong one (at the level of each sub-ensemble defined by the strong measurement outcomes), whether performed before or after them, to the extent that enables Alice to correctly identify which particle was subjected to which spin measurement by Bob, and what outcome was obtained*. The correlations between weak and strong outcomes are calculated in appendix 1.

b) Hence, all same-spin WMs confirm also one another.

c) Even the third spin orientation weakly measured by Alice, $\gamma$, is correlated with $\alpha$ and $\beta$ according to the same probabilistic relations of Eq. 1.

d) Even in case Bob's measurements are along orientations other than $\alpha$, $\beta$, or $\gamma$, Alice's data can precisely reveal these orientations, as well as all the individual spin outcomes, by employing the relation in Eq. 1.

These predictions are unique in two respects. The WM results robustly repeat themselves, *i.e.* give, after slicing, the same overall spin value. For example, the spin along the α-orientation remains the same upon the next spin α WM despite the intermediate β and γ VMs. Weak values, as emerging upon summation and proper normalization, thus seem to be unaffected by uncertainty relations, in contrast to the individual WM outcomes which seem almost random.

Even more striking is the fact that all WMs equally agree with the past and future strong measurements. While it is not surprising that the noon WMs *confirm* the morning strong outcomes, the equal degree of agreement with the evening ones can suggest that they *anticipate* them.

There is, of course, also a one-vector account for this result: The weak measurements introduce a slight bias (or "weak collapse") to each particle, which the later strong β measurements' outcomes may merely finalize, despite the other intermediate α and β weak measurements. It thus invokes two different explanations for the WM's agreement with the past α and future strong measurements. The past α value, so goes the one-vector account, has been already "collapsed," so the α WM only passively recorded it together with a great deal of noise, while the future β value was, *contrary to the philosophy of ABL*, still inexistent, hence the β WM exerted only a bias ("weak collapse") with respect to it, later to be overridden by the evening β measurement's results. TSVF, in contrast, simply asserts that both α and β values have coexisted during the intermediate interval between the two strong measurements. One may further point out that the correlation between the strong α and β is not precisely $cos^2\theta_{\alpha\beta}$ but slightly spoiled by a small bias of $\sim\lambda^2$ in each particle, indicating the disturbance brought about by the WMs.

We believe, however, that the experiment's results accord more with the ABL and the TSVF ontology. After all, *WM reveals just what TSVF predicts, a prediction that would never cross one's mind within standard quantum theory*, despite the equivalence between the two formalisms. Put differently, the WM's "bias", invoked above by the one-vector account, is viewed within the TSVF as affecting *both past and future* strong measurements, granting the measured particles well-defined properties (the weak values) in between the two strong measurements. Clearly the one-vector account lacks the TSVF's simplicity and elegance.

We will return to this comparison between the one- and two-vector accounts in Sec. 5.

## 4. Strong and Weak Measurements in the EPR-Bell Experiment

We can now demonstrate the weak outcomes' possible anticipation of a future human free choice. Consider an EPR-Bell experiment [1,2] on an ensemble of *N* particle pairs. Then,

### 4.1. Procedure

a) In the morning, Alice carries out three WMs on each particle within each pair, along the spin orientations $\alpha$, $\beta$ and $\gamma$. Every result is recorded together with all details: the pair's serial number among the *N*, the particle's identity (denoted by Right/Left for the two halves the entangled state) within the pair, and the weak measurement's number among the 3 (Fig. 2). The resulting 6 lists are then engraved on stone (as in Fig. 1) along 3 rows on each side.

b) In the evening, Bob, *oblivious of Alice's data*, performs one strong measurement on each particle. With sufficiently large *N* (enabling large enough sub-ensembles), he can choose a pair of measurements anew for each pair of particles. But for simplicity, let him choose one spin-orientation for all right-hand particles and one for all left-hand ones. The crucial fact is this: *The spin orientations are chosen at the last moment by Bob's free choice.*

c) Bob sends Alice his two binary lines, again not telling her whether "above/below line" refers to ↑/↓ or whether he has chosen α, β or γ for the right- and left-side particles.

d) Based on Bob's binary lines, Alice slices her data, carved on stone since morning, shifting each of the lines across her rows, as in Sec. 3.

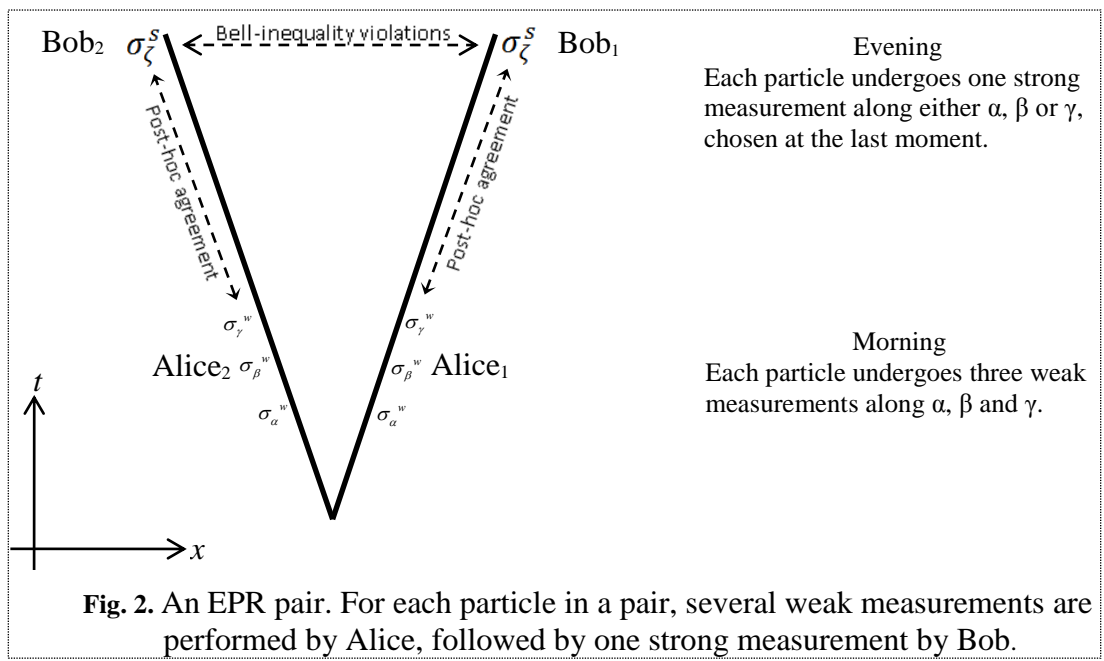

**Fig. 2.** An EPR pair. For each particle in a pair, several weak measurements are performed by Alice, followed by one strong measurement by Bob.

### 4.2. Predictions

Calculating the separate averages of each sub-ensemble, QM obliges the following (a statement about a WM refers to its overall outcome):

a) Bob's pairs of strong outcomes exhibit the familiar Bell-inequality violations [1], minus the slight $O(\lambda^2)$ spoiling [11] due to the intermediate WMs, indicating that the particles were mostly superposed prior to his measurements.

b) Alice's weak outcomes strictly agree with those of the strong measurements[7] at the level of each sub-ensemble;

c) with the following twist: The strong measurement of each spin determines also that of the other particle as if it has occurred *in its past*, *i.e.*, as in the experiments described in Sec. 3 with single particles, with the ↑/↓ sign inverted. This occurs regardless of the two measurements' actual timing [7].

## 5. What Kind of Causality?

The above experiment is designed to test the alleged time-symmetric quantum causality. What is the significance of the predicted results?

We begin with the main apparent contradiction. *i*) Bell's nonlocality proof states in effect that the particle pair could not possess pre-determined values. *ii*) In our experiment the WMs yield outcomes that turn out to match those of the strong measurements chosen *later*, thereby suggesting pre-existing weak values.

As TSVF and traditional quantum theory are equivalent, obliging one- and two-vector explanations to be equally valid, this contradiction can be resolved in two ways. Either *i*) the weak measurements have somehow anticipated the outcomes of the strong ones, chosen later (two-vector account), or *ii*) they have merely affected them in a subtle way (one-vector account). Obviously *ii*) is the more convenient option, especially with the alternative being retrocausal in character. The one-vector account, however, also comes with heavy conceptual cost:

---

[7] One could make the results even more dramatic by adding here another feature used in Sec.3, namely repeating the three weak measurements three times, revealing their remarkable robustness.

a) It relies on "weak biases" like those invoked in Sec. 3 for the single-particle case, whereas TSVF simply asserts that both measurements equally affect both particles, as in 4.2(c).

b) In time-symmetric QM, nonlocality turns out to be only apparent restricted to the three spatial dimensions, as in the 4-dimensional spacetime the quantum correlations are local. Indeed the EPR correlations are precisely such that the spin value of each particle is tantamount to a pre-selection of the other.

c) The mathematical simplicity and conceptual elegancy of the TSVF are evident. When the pre- and post-selected states are set, we immediately know what would be all the weak values of this sub-ensemble, without the need of calculating the consecutive non-commuting biases.

d) Moreover, not only were the above results predicted by TSVF, they are hardly likely to even cross one's mind in any one-vector formalism.

e) Additionally, anticipation of future outcomes is by no means the only surprise predicted by TSVF for this case. Recall first that, for mere convenience, we have treated all WM outcomes as binary, i.e., either ↑ or ↓, while in reality they can have various magnitudes, many of them unrealistic because of the strong noise. With a slight modification of the spin directions chosen to be measured for the EPR pair, and under appropriate post-selection, TSVF predicts that the particles will possess even more surprising outcomes [16-19]. Again, these predictions, while consistent with standard quantum theory, are extremely unlikely to have emerged within the standard formalism.

Alternatively, one can turn to the Many-Worlds Interpretation, where there is no "action at a distance" [20], but then the principle of simplicity is severely harmed. Finally, regardless of the one- or two-vector account one opts for, the following offshoots of our experiment merit attention:

1. **Counterfactuals actualized.** Counterfactuals seem to be part and parcel of the EPR setting, such as "Have the right-hand particle been subject to the measurement chosen for the left-hand one and *vice versa*, they would give the same spin values with the signs inverted." With WMs, these are no more counterfactuals, but actual outcomes. Moreover, our experiment turns even the more abstract counterfactual part of the EPR experiment into an actual physical result: Prediction (c) in Sec. 4.2 refers to a spin-orientation *not* eventually chosen for strong measurements, a counterfactual even farther from reality than the exchange between actual measurements. Yet in our setting, *even this unperformed choice yields actual results through the weak measurements*. This point is further discussed in [10,11]. Moreover, in [16,17] we discuss the issue of "odd" weak values (those that exceed the spectrum of the measured operator) which emerge naturally in these setups and lend further support to the TSVF.

2. **Subtle collapse:** Under a strict one-vector account, our experiment suggests a subtle way to bypass Bell's restriction on pre-existing values. The particles are allowed to possess *weak* values, enabling subtle methods of contamination of the results that invite further exploration.

3. **Subtler inequalities?** By being continuous rather than discrete, the weak pointer outcomes obtained by Alice, call for a new kind of Bell-like inequality.

## 6. Summary

We explored an apparent contradiction between two well-established findings:

a) The EPR-Bell experiment proves that one particle's spin outcome depends on the choice of the spin-orientation to be measured on the other particle, and its outcome thereof. Relativistic locality is not violated in this experiment due to the reciprocity between the two measurements, allowing either Alice's choices to affect Bob's, or vice versa.

b) Such reciprocity, however, is challenged for a combination of measurements of which one is strong and the other weak. The latter affects the former to a much lesser extent, i.e. all the weak outcomes do not oblige the strong ones, but the strong outcomes do determine the weak values. Moreover, the strong correlation between past and future outcomes, suggests the appearance of a subtle local hidden variable – the future state vector.

Therefore, when a weak measurement precedes a strong one within an EPR experiment, the weak-strong agreements between past and future measurement outcomes can be interpreted in two ways. One may adhere to the one-vector non-local explanation and ascribe it to the slight biasing of the weak on the strong measurements. Simplicity and elegance, we suggest, favors the local two-vector account, where the future choice plays a crucial role within the noisy weak outcomes carved on the rock.

We currently collaborate with the team of Prof. Genovese, to perform an optical variant of this gedanken experiment in the laboratories of the Italian National Institute for the Research of Metrology (INRIM) at Torino. Results will be published in due course.

## Acknowledgements

It is a pleasure to Yuval Gefen, Nicolas Gisin, Doron Grossman, Ruth Kastner and Marius Usher for helpful comments and discussions. This work has been supported in part by the Israel Science Foundation Grant No. 1311/14.

## Appendix 1. The Objective Nature of Quantum Information

When a sequence of signals turns out to carry some information, such as radio signals giving a picture of a distant galaxy on a radio-telescope's screen, it is obvious that the information is encoded within the *signals* rather than within their code or decoding method. The TSVF makes a similar argument about Alice's weak measurements' outcomes: *They* already contained the information about Bob's future choice, even before he himself knew what he will decide.

Let us formulate this argument mathematically. Decoding an encoded message means finding the "vertical" correlations between the message's signals and those of the original script, *e.g.*, a=b. Yet there are also "horizontal" correlations within the original script's signals, *e.g.*, in English, the greater probability for "qu" juxtaposition or the greater abundance of "m" over "z". Detecting such horizontal correlations within the coded message facilitates revealing the vertical correlations with the script.

Let us apply this method to the EPR experiment. Here too, the nonlocal effect is revealed only after each list of outcomes is sliced in accordance with the other. To stress the effect's objectivity, consider the following Alice-Bob variant. Bob is using a Chinese spin-measuring apparatus. Not knowing which spin-orientation he measures, neither the spin value he gets, he sends Alice a list with only his outcomes' division into the groups *xI, xII, yI, yII, zI* and *zII*, denoting the three unknown spin orientations and their two possible values. Alice, slicing her data into the corresponding sub-ensembles, reveals the horizontal correlations within her data, which, by their correspondence with the horizontal correlations within Bob's data, yield all the vertical correlations she needs. She can now provide Bob with the precise *α/β/γ/↑/↓* value for each of his own measurements. Ergo, the

EPR nonlocal effect resides within the measurements' data, independently of the code for its slicing.

The underlying mathematical principles are simple. Let $A_1,...,A_n$ be Alice's and $B_1,...,B_n$ Bob's outcomes, respectively, along a certain direction. The "vertical" correlations between their lists are given by the Pearson Coefficient:

$$r_{AB} = \frac{\sum_{i=1}^{n}(A_i - \bar{A})(B_i - \bar{B})}{(n-1)s_A s_B}, \qquad (9)$$

where $S_A$ and $S_B$ are the sample horizontal standard deviations of $A_i$ and $B_i$. $B_i$ are either $1/2$ or $-1/2$ with 0 average, so

$$s_B = \frac{1}{2}\sqrt{\frac{n}{n-1}}. \qquad (10)$$

Next we express the nominator of Eq. 9 in a more explicit way:

$$r_{AB} = 2\frac{\sum_{i=1}^{n}(\frac{1}{2}\frac{\lambda}{\sqrt{n}}(-1)^{a_i} + \Delta_i)(\frac{1}{2}(-1)^{b_i})}{s_A\sqrt{n(n-1)}}. \qquad (11)$$

Since $a_i=b_i$ for every $i$ and the noise $\Delta_i$ is symmetric,

$$r_{AB} = \frac{\lambda}{2s_A}\sqrt{\frac{1}{n-1}}. \qquad (12)$$

This correlation is much higher than the correlation which a 1-vector account would have predicted (i.e. without the post-selection information), because $a_i$ and $b_i$ can be different. According to orthodox quantum mechanics, there is a chance, say $0.5+\varepsilon$ (which depends on the strength of the weak measurement) to have $a_i=b_i$, and therefore, a factor of $2\varepsilon$ enters Eq. 11.

The 1-vector account of this result invokes mere "leakage" of information from the original "text" namely the weak outcomes, to the "code" i.e., the strong ones. The pros and cons for the two options were discussed in the concluding sections.

# Appendix 2. On the emergence of the TSVF from orthodox quantum mechanics

Quantum mechanics states that a system prepared at $|\psi_{in}\rangle$ will propagate towards $|\psi(t)\rangle = U(t_{in},t)|\psi_{in}\rangle = \exp[-iH(t-t_{in})/\hbar]|\psi_{in}\rangle$ at time $t > t_{in}$ and would collapse to the eigenstate $|a_j\rangle$ of an operator $A$ with probability $|\langle a_j|U(t_{in},t)|\psi_{in}\rangle|^2$ upon being measured.

For $t_{fin} > t$ The connection between future and past time-evolutions is:

$$U(t_{fin},t) = U^{-1}(t,t_{fin}) = U^\dagger(t,t_{fin}).$$

Therefore, as in the forward time evolution, the final state of the system $|\psi_{fin}\rangle$ propagates towards the past with $\langle\psi(t)| = \langle\psi_{fin}|U(t,t_{fin})$ and will coincide with $|a_j\rangle$ with probability $|\langle\psi_{fin}|U(t,t_{fin})|a_j\rangle|^2$.

According to Bayes rule:

$$\Pr(aj,t\,|\,\psi_{in},t_{in},\psi_{fin}t_{fin}) = \frac{|\langle\psi_{fin}|U(t,t_{fin})|a_j\rangle\langle a_j|U(t_{in},t)|\psi_{in}\rangle|^2}{\sum_i |\langle\psi_{fin}|U(t,t_{fin})|a_i\rangle\langle a_i|U(t_{in},t)|\psi_{in}\rangle|^2}. \tag{13}$$

For this reason, the TSVF stems directly from orthodox quantum mechanics, but in order to explain the former's predictions, the latter would have to either employ a future state vector or resort to miraculous conspiracies between *errors*.

# Appendix 3. Weak Measurements in the POVM Formalism

It might be instructive to use the POVM formalism for denoting "weak" measurements and compare them with "strong" measurements.

**1. "Strong" measurements:** Suppose a measurement described by a set of operators $\{M_m\}_{m=1}^n$ is performed upon a quantum system prepared at $|\psi\rangle$. Then the probability for outcome $m$ is $p(m) = \langle \psi | M_m^\dagger M_m | \psi \rangle$ [21]. If the quantum state is given in terms of a density matrix $\rho_i$, then $p(m) = Tr\{M_m \rho_i M_m^\dagger\}$, and the joint probability of obtaining first $m$ and then $f$ is given by $p(m, f | i) = Tr\{\Pi_f M_m \rho_i M_m^\dagger\}$, where $\Pi_f$ is another POVM [22]. The POVM elements are defined by the projectors $E_m = M_m^\dagger M_m$, $m = 1,...,n$. For every $m$, $E_m$ is a positive operator such that: $\sum_{m=1}^n E_m = I$ and $p(m) = \langle \psi | E_m | \psi \rangle$. Thus the set of operators $E_m$ is sufficient to determine the probabilities of the different measurement outcomes.

**2. "Weak" measurements:** The weakness of the measurement is quantified by the closeness of the measurement operators $M_m$ to multiples of the identity operator $I$. This can be done by the following representation of each $E_m$ [21]: $E_m = w_m(I + \varepsilon S_m)$, where: $\sum_{m=1}^n w_m = \frac{1}{1+\varepsilon}$ and $\sum_{m=1}^n w_m S_m = \frac{I}{1+\varepsilon}$. The measurement probabilities can now be expressed in terms of the expectation values of a set of self-adjoined operators $S_m$ that is independent of the measurement strength $\varepsilon$. Furthermore, in the limit of $\varepsilon \ll 1$ the measurement operators $M_m$ are approximately given by the linearized square root of the POVM operators

$M_m \approx \sqrt{w_m}(I + \frac{\varepsilon}{2}S_m)$. In addition, the above joint probability can be approximately expressed by: $p(m,f|i) \approx Tr\{\Pi_f \frac{1}{2}(\rho_i M_m^\dagger M_m + M_m^\dagger M_m \rho_i)\}$.

The important point is the final measurement probabilities $p(f|i)$ are not changed by the measurement of *m, i.e.* $p(f|i) = \sum_{m=1}^{n} p(m,f|i) = Tr\{\Pi_f \rho_i\}$.

In our scheme (Sec. 4) Alice performs 9 such measurements on each particle of the EPR pair. Later, Bob measures strongly each particle of the "weakly collapsed" pair. When dividing her weak outcomes according to Bob's outcomes, Alice finds significant past-future correlations, at the level of each sub-ensemble. This procedure enables Alice to assign (in retrospect) a weak value to each of her particles.

The above method for performing weak measurements is possible due to Naimark's dilation theorem which asserts that any POVM can always be realized with a suitable coupling between the measured system and an auxiliary system (an ancilla) by performing a projective measurement on the joint system, provided the auxiliary system is large enough [23]. Our gedanken experiment strongly relies on this fact.